# Rogue waters

Alexey Slunyaev (corresponding author)

*1) Department of Nonlinear Geophysical Processes, Institute of Applied Physics, Nizhny Novgorod, Russia*

46 Uljanov Street, Nizhny Novgorod, 603950 Russia
Phone: +7 831 4164674,    Fax: +7 831 4365976
slunyaev@hydro.appl.sci-nnov.ru

*2) Department of Mathematics, Research Institute for the Environment, Physical Sciences and Applied Mathematics, Keele University, Keele UK*

EPSAM, Keele University, Staffordshire, ST5 5BG, UK
Phone: +44 1782 733414,    Fax. +44 1782 584268

Ira Didenkulova

*1) Department of Mechanics and Applied Mathematics, Institute of Cybernetics, Tallinn, Estonia*

Akadeemia tee 21, 12618 Tallinn, Estonia
Phone: +372 6204260,    Fax: +372 6204151
ira@cs.ioc.ee

*2) Department of Nonlinear Geophysical Processes, Institute of Applied Physics, Nizhny Novgorod, Russia*

46 Uljanov Street, Nizhny Novgorod, 603950 Russia
Phone: +7 831 4164674,    Fax: +7 831 4365976

Efim Pelinovsky

*1) Department of Nonlinear Geophysical Processes, Institute of Applied Physics, Nizhny Novgorod, Russia*

46 Uljanov Street, Nizhny Novgorod, 603950 Russia
Phone: +7 831 4164839,    Fax: +7 831 4365976
pelinovsky@hydro.appl.sci-nnov.ru

*2) Department of Information Systems, Higher School of Economics, Nizhny Novgorod, Russia*

24 Pecherskaya Street, Nizhny Novgorod, Russia
Phone: +7 831 4169549
epelinovskiy@hse.ru

A. Slunyaev received his PhD in Mechanics from the Mediterranean University in 2002 and Cand. Phys. Math. Sci. Degree in the Institute of Applied Physics the same year. He was awarded the Medal of the Russian Academy of Sciences for young researchers in 2006 for the study on rogue wave phenomenon. He is permanently affiliated by the Institute of Applied Physics. In 2010 his research was supported by the Marie Curie fellowship grant.

I. Didenkulova received her Cand. Phys. Math. Sci. Degree in Fluid Mechanics (2006) and PhD in Civil Engineering (2008). Her research is devoted to the complex behaviour of different marine natural hazards, including tsunamis, freak waves, storm surges in the coastal zone. She was awarded the Plinius Medal by the European Geosciences Union in 2010 in recognition of her contribution to solve complex problems of oceanography and coastal engineering.

E. Pelinovsky is an international expert in the nonlinear wave physics and its applications to marine natural hazards. He received his Cand. Phys. Math. Sci. Degree in 1972 and Dr. Phys. Math. Sci. Degree (the highest scientific degree in Russia) in 1981. He was awarded the Russian State Award in 1996 and the Soloviev Medal by the European Geosciences Union in 2006. He is the author of nine books including two published in English: "The nonlinear evolution equations" (Longman, 1988) and "Rogue Waves in the Ocean" (Springer, 2009).

# Rogue waters

In this essay we give an overview on the problem of rogue or freak wave formation in the ocean. The matter of the phenomenon is a sporadic occurrence of unexpectedly high waves on the sea surface. These waves cause serious danger for sailing and sea use. A number of huge wave accidents resulted in damages, ship losses and people injuries and deaths are known. Now marine researchers do believe that these waves belong to a specific kind of sea waves, not taken into account by conventional models for sea wind waves. This paper addresses to the nature of the rogue wave problem from the general viewpoint based on the wave process ideas. We start introducing some primitive elements of sea wave physics with the purpose to pave the way for the further discussion. We discuss linear physical mechanisms which are responsible for high wave formation, at first. Then, we proceed with description of different sea conditions, starting from the open deep sea, and approaching the sea cost. Nonlinear effects which are able to cause rogue waves are emphasised. In conclusion we briefly discuss the generality of the physical mechanisms suggested for the rogue wave explanation; they are valid for rogue wave phenomena in other media such as solid matters, superconductors, plasmas and nonlinear optics.

## Outline

# 1. Introduction. Physical mechanisms

The popularity of sea freak or rogue wave problem amplified suddenly within the last few decades, and it is now a top-rank topic for scientific discussions, conferences and publications. The adequate estimate of the rogue wave frequency, kinematics, dynamics, and the rogue wave danger forecasting are of a great economical importance. This vogue was boosted both by theoreticians who managed to reveal new extraordinary features of the nonlinear wave dynamics, and marine engineers thanks to trustworthy testimonies of the ultimate effects caused by the extreme waves. The rogue wave phenomenon and related problems are described in a very popular way in [Lawton, 2001; Garrett & Gemmrich, 2009; Ridgway, 2010]. Scientific reviews on the rogue wave problem in the physical oceanography may be found in [Kharif & Pelinovsky, 2003; Dysthe et al, 2008; Kharif et al, 2009], collections of research papers are gathered in proceedings of topical conferences [Olagnon & Athanassoulis, 2001; Olagnon & Prevosto, 2005, 2009]. A more general view on rogue waves in physics and mathematics is given in the special issue of the European Physical Journal [Akhmediev & Pelinovsky, 2010].

It seems so that the term *freak wave* was more popular at the beginning of the story, when the phenomenon of sudden and unexpectedly high waves occurring in the sea had become recognized by the scientific community. It was accepted that these waves were something beyond *extreme waves* or *steep wave events*.

The marine folklore could enrich the physical vocabulary with many colourful names for the dangerous oceanic waves, such as abnormal, exceptional, extreme, giant, huge, sudden, episodic, monster, vicious, killer, mad- or rabid-dog waves; cape rollers, holes in the sea, walls of water, three sisters… Now people seem to prefer calling these waves *rogue waves* instead of freak waves, what probably reflects some level of understanding of this phenomenon, and the sign that the problem is now becoming more developed, though still dangerous and challenging. In this paper an introduction to the present understanding of the rogue wave phenomenon is given.

Usual waves on the sea surface are generated by winds. Their dynamics is strongly affected by weather conditions and oceanographic conditions (bathymetry, currents), which may be both varying. The variability of the conditions and co-existence of many wave systems generated by different winds at different water areas result in very complicated *stochastic* dynamics of sea waves. Revealing relations between weather conditions and dangerous for navigating sea states is an indispensable need for providing the safe sea use. The on-going research is eventually directed towards solution of this vital problem.

Sea waves are a continual object for studying. A significant number of surface elevation time series records containing rogue events may be found in literature. In-situ measurements are exploited with the purpose to reveal the physical origin of the rogue wave effect, to determine most dangerous sea states, and to find a reliable indicator of the high probability of rogue wave occurrence. Rogue wave records are also used for reconstruction of the dangerous events, initializing numerical simulations and verifying their abilities. These issues have been addressed by many researches, and are discussed in the book [Kharif et al, 2009]; though the present paper is dedicated to a more general overview on physical mechanisms of rogue wave formation.

Though large waves are usually expected to face at severe sea states, rogue waves are observed at calm sea conditions as well. This confirms the significance of the *own sea wave dynamics*, at which we restrict our attention in this communication. The interaction between intense waves and winds represents a more difficult problem which is not sufficiently investigated at the present moment. Some information on these recent studies may be found in [Kharif et al, 2009]. Since the wind-wave interaction is much slower than the wave-wave interaction, in many cases rogue waves may be supposed free from the wind action. The wind effect contribution to the process of rogue wave formation will not be considered in the paper.

The following main physical mechanisms of rogue waves may be conventionally singled out:

**Geometrical focusing** (spatial focusing). This effect is well-known in physics, and particularly in optics. The focusing of waves coming from different directions may be caused by the interference of different sea wave systems, generated at different storm areas (*mixed sea* states), or/and by wave refraction and diffraction by bathymetric peculiarities (underwater hills, ridges, etc), and by oceanic non-uniform currents as well. The spatial focusing can result in significant wave amplification at regions of caustics (wave focuses). The unexpectedness of huge wave formation, commonly attributed to the rogue wave nature, is due to the variability of wind wave patterns in storm areas, when a weak variation of wave paths can lead to the disappearance of existing caustics and their formation elsewhere, likely far from the previous place.

**Focusing due to dispersion of the wave group velocity** (temporal focusing). Water waves are dispersive waves as far as different spectral components propagate with their own velocities. During wave propagation, components of various scales may merge at a single place resulting in a strong energy concentration (dispersive focus). The process of wave dispersive focusing persistently occurs for ocean waves; the formation of intense waves has a random character due to the variability of wind wave patterns in storm areas.

**Focusing due to the modulational (Benjamin – Feir) instability**. This effect is due to the modulational instability of nonlinear waves, which is also widely known for waves of different physical origins; see historical essay [Zakharov & Ostrovsky, 2009]. In water wave physics it is called the Benjamin – Feir instability after T.B. Benjamin and J.E. Feir who discovered this effect in a laboratory tank in 1967. The essence of this phenomenon is in an unstable growth of weak wave modulations, which evolve into short groups of steep waves. Through this dynamics, the wave energy gets focused for a short time, providing a rogue character of the modulational instability effect.

Nonlinear water waves suffer from different kinds of nonlinear instabilities depending on the wave intensity and the water depth, which may modify the action of the Benjamin – Feir instability or result in similar effects. A proximity to homoclinic orbits of the nonlinear evolution equations, governing the wave dynamics, is suggested as the explanation of the freaky dynamics of rogue waves (see [Calini & Schober, 2002] and book [Osborne, 2010]).

**Essentially nonlinear wave interaction.** Nonlinear effects, when waves interact at certain conditions, are able to produce much more significant wave amplification than it is expected from the linear superposition assumption. This effect strongly depends on the angle between the waves, wave shapes (soliton or shock waves), nonlinear characteristics, and so on, providing the rogue character of such interaction.

**Wave-current interaction**. This source of rogue waves is historically singled out, because one of the first notorious areas where the rogue wave effect was recognized were waters with strong currents (for example, the Agulhas current off the Southeast coast of Africa). Effects of wave trapping and wave blocking are explained by the strong influence of the opposite current upon the wave dispersion low. From this point of view the wave propagation over currents and wave blocking are in some sense similar to the wave refraction and wave reflection at the shallow coastal waters; these effects have become classic, and will not be analyzed in detail in the present paper.

The most of the listed effects have general physical matter and potential applications to many other fields of physics.

To start, in the next section we introduce the main definitions and features of sea waves, which will be relevant for the further understanding. Proceeding from the simple theory to a more advanced, we first consider linear effects, which may cause an extreme wave generation, and then proceed to nonlinear effects. The recent achievements in understanding of the rogue wave phenomenon are mostly related to sea wave nonlinear effects, and the present paper emphasizes their contribution. Regarding the features of wave dynamics conditions, the subsequent sections describe rogue waves over deep and shallow waters. In the conclusion some closely related problems in physics are briefly discussed.

**2. General issues**
In this section some general properties of surface wind gravity sea waves are described to prepare for the further reading.

*2.1. Wind waves*
Waves with length within the range of about 10 cm – 500 m, which are usually observed on the sea surface, are *wind-generated* waves. They are a part of the class of water waves, *surface gravity waves*, propagating due to the action of gravity. Far from the storm area degenerating wind waves become longer and less steep, and are called *swells*; they can influence and interact with the local wind waves and can result in the rogue event. .

On the wind wave crests short *capillary waves* are usually generated (of a few centimeter length and shorter); their dynamics is governed by the surface tension at the boundary between the water and the air. These waves are important for processes of the ocean-atmosphere interaction, but they do not contribute to the rogue wave phenomenon.

### 2.2. *Deep and shallow water waves*

Surface gravity waves obey the dispersion law

$$\omega = \sqrt{gk \tanh(kh)}, \qquad (1)$$

where $\omega = 2\pi/T$ is the cyclic wave frequency ($T$ is the wave period), $k = 2\pi/\lambda$ is the wavenumber ($\lambda$ is the wavelength), $g$ is the gravity acceleration, and $h$ is the water depth.

Sinusoidal waves propagate with the phase velocity, $C_{ph} = \omega/k$, though wave groups travel with the group velocity, $C_{gr} = d\omega/dk$. The wave energy is transferred with the group velocity as well.

Water waves propagating on the sea surface induce the motion of fluid particles within the water column, which depends on the ratio between the wavelength and the water depth. Hereafter, the deep and shallow water conditions will be divided in relation to the surface wave physics. If $kh \gg 1$, surface waves induce fluid motions which decay with depth exponentially and the propagating wave is not influenced by the bottom; hence, the sea is assumed deep (*deep water waves* or *short gravity waves*). The deep-water dispersion relation has the form

$$\omega = \sqrt{gk}, \quad C_{ph} = \sqrt{\frac{g}{k}} = \frac{g}{\omega}, \quad C_{gr} = \frac{1}{2}\sqrt{\frac{g}{k}} = \frac{g}{2\omega}. \quad (2)$$

Therefore, individual waves in a group move twice faster than the wave group as a whole. Due to the difference between the phase and group velocities, individual waves are enclosed within the wave group: they appear at the rear part of the group, run to the group frontal part and seem disappearing.

In contrast to the deep-sea conditions, *shallow water waves* ($kh \ll 1$) represent the wave motion involving the entire water column, and the water flow is almost uniform within the depth. The dispersion relation for shallow water waves gives

$$\omega \approx Ck\left(1 - \frac{k^2 h^2}{6}\right), \quad C_{ph} \approx C\left(1 - \frac{k^2 h^2}{6}\right), \quad C_{gr} \approx C\left(1 - \frac{k^2 h^2}{2}\right), \quad C \equiv \sqrt{gh}. \quad (3)$$

Shallow water waves possess a weak difference between phase and group velocities, and are weakly dispersive waves. As a result, individual waves can propagate over a long distance without significant transformation; that is used by windsurfers near the coast.

Tsunami wave crossing the ocean is another bright example of shallow-water waves. A general review on effects of tsunami wave generation and propagation can be found in [Levin & Nosov, 2008]. Tsunamis can result in occurrence of huge waves (30-40 m at the coast and up to several meters in the open sea), and two latest global events (2004 Indian Ocean tsunami, and 2011 Japanese tsunami) are well described in mass-media. Meanwhile, tsunami waves are not rogue (unpredictable) waves, and, therefore, are not in the focus of the present paper.

The sea wave nonlinearity is evidently manifested through the sharp wave shape. Indeed, only small-amplitude or long waves are close to sinusoidal. Steeper waves are asymmetric; too much steep waves cannot propagate and break. In fact, the nonlinearity has two main consequences: the appearance of phase-locked harmonics and the energy exchange between wave harmonics. Both effects violate the Gaussian sea assumption which will be discussed shortly below.

The presence of phase-locked spectral components results in the deviation of the wave shape from sinusoidal. The mutual dependence of wave harmonics leads to the formation of coherent states; the energy exchange between Fourier modes may be unstable and cause transfer energy to other scales, and as a result, may cause the occurrence of short intense wave packets and steep waves. Besides, nonlinearity means cancellation of the linear superposition assumption, what complicates the wave description.

Fig. 1 displays recorded wave profiles when waves approach the coast: the deep-water region corresponds to Fig. 1a; Fig. 1b reports on the shallow-water condition [Cherneva & Guedes Soares, 2005]. The difference between the wave appearances is obvious. Waves over shallow water are much more asymmetric and have high steep crests. This difference is caused by the nonlinearity, which acts differently for deep and shallow waters.

**2.3.** *What a rogue wave is*

Although the problem of rogue waves has been studied for a few decades, a single generally accepted definition of a rogue wave still does not exist. It is commonly assumed that these waves should be defiantly high and ruinous, and also unexpected.

By now there is a number of well-documented cases of occurrence of unexpectedly large sea waves. When talking about the frequency of high waves, the key factor is the balance between the wave size and its probability. It is that very condition which is incorporated into the ship and marine structure design rules, and risk assessments. While the risk to be hit by an extraordinary high wave may be negligible, the chance to face moderately high waves can be quite probable and must be taken into account instead.

The wave height, $H$, is the most evident quantitative estimate of the wave size. It is defined as the vertical distance between the wave crest and the deepest trough preceding or following the crest. Frequently, a simple definition of a rogue wave is employed, that this wave at least twice exceeds the significant wave height:

$$AI > 2, \quad \text{where} \quad AI = \frac{H_{\max}}{H_s}. \qquad (4)$$

Here $H_{\max}$ is the height of the rogue wave, and $H_s$ is the *significant wave height*[1] which is the averaged of one third highest waves in a time series (usually the time series has duration 10–30 min, what corresponds to about 50-300 individual waves).

The wave height is not the only significant injurious factor that makes waves rogue. The wave impact upon marine stationary or moving structures may be governed by other parameters, such as steepness (which is proportional to $H_{\max} / \lambda$, where $\lambda$ is the wave length), crest height, horizontal and vertical wave asymmetry; specific wave sequences are also expected to be quite dangerous due to the hull memory and resonance effects. Different types of ships may also suffer from different wave parameters and sea conditions. Therefore, different kinds of extreme waves may be treated as rogue waves or not, depending on the particular case and applications.

The rogue wave impact is an important practical question. However in this paper we define a rogue wave following the only amplitude criterion on the 'abnormality index', *AI*, given above.

## 2.4. What the maximum attainable sea wave height is

When in 1826 Captain Dumont d'Urville, a French scientist and naval officer in command of an expedition, reported encountering waves up to 30 meters height, he was openly ridiculed. Three of his colleagues supported his estimate but could not help him to be admitted [Draper, 1964]. According to modern reliable instrumental measurements, wind waves can have a height of 30 m, they have been repeatedly registered during storms. As an example, waves with heights of a little bit more than 29 m were measured under severe but not exceptional wind conditions in 2000 by a British oceanographic research vessel near Rockall, west of Scotland. A three-week registration of surface waves from the European satellite ERS-2 revealed regions in the World Ocean with high waves and detected a wave of 29.8 m height. Even higher waves have been reported, but those testimonies seem to be doubtful, see details of rogue wave observations in [Kharif et al, 2009] and references therein.

---

[1] This is a specific sea term; it equals to the fourfold covariance, $H_s \approx 4\ \sigma$, if the waves represent a random Gaussian process.

Bearing in mind that ships are often designed for 10–15 m wave heights, it becomes obvious that the observed waves are real threats that may cause a significant damage and even a ship loss.

## 2.5. Rogue wave statistics

The concept of a Gaussian random process is very popular in physics, and the *Gaussian sea* assumption is the most conventional one when wind waves on the sea surface are concerned. In this case surface waves are considered to be a linear superposition of many independent harmonics[1], which propagate in different directions with different speeds and make the sea surface movement *stochastic*.

If so, then the Central Limit Theorem insures that the sea surface displacement obeys the normal (Gaussian) distribution. Under the assumption of a narrow banded wave spectrum (a rough approximation of the real wind wave spectrum) it results in the Rayleigh distribution function for the wave height probability, so that the exceedance probability function is

$$P(H) \approx \exp\left(-2\frac{H^2}{H_s}\right), \qquad (5)$$

see, for example [Massel, 1996]. It may be straightforwardly obtained that this estimate foresees the formation of a rogue wave with $AI > 2$ at a single measuring point every 8–9 hours for the typical wave period of 10 s. A wave satisfying the condition $AI > 3$ would be measured once in about 20 years, and such waves have already been reported.

---

[1] The Fourier spectrum of wind waves is often supposed to be concentrated around the central frequency, which value is determined by the wind action, and the spectrum width is less or comparable with the central frequency value.

Theoretically, the Rayleigh distribution function means that a wave of any height may occur. In practice, high waves are influenced by nonlinearity, and, thus, the tail of this distribution is different. The true shape of the tail of the wave height exceedance probability function is the cornerstone, which determines the importance of the rogue wave problem. According to some in-situ observations, at the values of the attainable wave amplification $AI \approx 2 \ldots 4$ the wave height probability many times exceeds the Rayleigh distribution.

People have started instrumental recording of sea waves since the middle of the 20th century. High waves are of the prior interest, and, thus, most of the wave measurements are acquired during the stormy weather. Technical limitations make the registration of extreme waves even more difficult, what complicated assembling of rich statistical data on rogue wave events. To the best of our knowledge, the number of recorded rogue waves throughout the world accounts a few thousands, but the percentage of trustworthy ones is much smaller.

The recorded waves belong to different sea states and do not compose a statistically uniform ensemble. This prevents obtaining the reliable rogue wave statistics on the basis of natural records, and, hence, the rogue wave probability problem is addressed to theoretical analysis and wave simulations.

**3. Linear rogue waves**

It is obvious that in the linear theory rogue waves can be formed due to focusing mechanisms only. These mechanisms are manifested differently for deep and shallow waters.

In the *shallow water* due to the strong influence of the variable bottom relief, the spatial focusing of waves is the major mechanism of rogue wave generation. Inhomogeneous depth conditions lead to varying wave speeds, and effects of refraction result in concentration of wave energy at caustics. Fig. 2 demonstrates ray patterns for shallow water waves generated by an isotropic source in the Japan Sea.

Any weak variation of the wave source (such as a storm area) parameters dramatically modifies the location and intensity of focal spots, what may explain the rapid appearance and disappearance of intense wave areas and related extreme waves. Other factors which alter and contribute to the wave focusing are: a wave-current interaction, a superposition with waves reflected from the coast, resonance effects in closed and semi-closed basins, and diffraction effects. The matter of rogue waves in all these cases is common, and the unexpected nature of rogue waves is supported by the random location of focal points and its sensitivity with respect to weak variations of the source. The effect of temporal focusing due to the weak dispersion can also be important for waves in long channels of almost constant depth.

In the *deep water* the effect of geometrical focusing is significant in relation to wave-current and wave-swell interactions (crossing seas), to the non-uniformity of the storm area with several centers of wave generation; but the effect of temporal focusing is important in all cases.

In Fig. 3 the local wave group velocity computed for a 20-min time series is shown. A significant variation of the wave velocity is evident. It is the reason for continual dispersive wave focusing and defocusing that leads to the occasional formation of larger waves.

Supposing the waves to be linearly superposing, tailored wave sequences can be prepared to provide the wave focusing at a single point at one time instant. The corresponding wave condition may be easily found with the use of the dispersion relation formula. This approach turns out to be convenient for generation of large-amplitude waves in laboratory conditions, see Fig. 4 for example. The wave profiles at different distances from the wavemaker are shown on the right. It may be seen that the initial wave packet contains waves with different lengths and, respectively, periods (a frequency modulated wave train). The difference in wave length provides the condition for efficient wave focusing downstream the wave tank due to the dispersion. The focused wave train contains a single wave oscillation, which breaks shortly after it is formed.

Waves generated in a laboratory tank are naturally influenced by perturbations due to the imperfectness of the equipment, and also by nonlinear effects. In spite of this, transient waves are successfully used to produce dispersion-generated extreme waves in flumes. This fact demonstrates the ability of the dispersive focusing effect to act in the real conditions, and confirms its robustness.

The effect of sea wave nonlinearity is now believed to be the prevalent mechanism of rogue wave probability increase beyond the conventional linear (or quasi-linear) theories. Extreme wave impact on ships and marine structures is also associated with strongly nonlinear effects. Nonlinear corrections to the wave shape and kinematics, and essentially nonlinear phenomena in the wave dynamics, all have effect on the rogue wave and extreme load probabilities. This vital problem has motivated much study on the nonlinear rogue wave dynamics and statistics.

**4. Nonlinear deep-water rogue waves**

An example of the surface elevation time series retrieved on the Draupner stationary platform, situated in the North Sea, is given in Fig. 5. The upper panel shows the 20-min record of the famous New Year Wave recorded on Jan 01, 1995. The lower panel represents it in a larger scale. Circles over the curves in the lower panel denote the measured values. This extraordinary wave had hit the platform and, hence, excited interest to the rogue wave problem greatly.

In general, different kinds of rogue wave appearance may be specified: single waves and wave groups (sometimes called "three sisters"), pyramidal waves and walls of water, etc. So-called 'holes in the sea', which are very deep wave troughs, are also observed, and they are frequently supposed to be even more dangerous than huge wave crests, because the deep trough cannot be seen when hidden behind surrounding waves.

In this section the peculiarities of the nonlinear mechanisms leading to the formation of rogue waves in deep waters are discussed.

*4.1. Stokes waves*

First of all, nonlinearity influences the wave shape and the speed of wave propagation. The basic example of deep-water nonlinear waves is the travelling waves with a permanent shape, which are called *Stokes waves*. Their crests are sharp, and their troughs are smooth, see Fig. 6. The degree of the wave nonlinearity regardless the water depth may be measured in terms of a dimensionless parameter, the wave steepness, $s = kH/2$. Uniform waves over the deep water (the Stokes waves) break when the wave steepness is about $s \approx 0.4$. Typical intense sea wave trains are characterized by the steepness of about $s \approx 0.07…0.1$.

It is well-known that the nonlinearity of Stokes waves results in increase of the wave frequency. The deep-water nonlinear frequency correction is proportional to the squared wave amplitude, $|A|$,

$$\omega = \omega_{linear}\left(1 + \frac{1}{2}|kA|^2\right), \qquad (6)$$

where $\omega_{linear}$ is the frequency of linear waves, see Eq. (2). The nonlinear frequency correction leads to important effects in the nonlinear wave-wave energy exchange. Waves with uniform lengths may propagate with different velocities due to the amplitude modulation, what gives a possibility for a nonlinear wave focusing. This effect becomes apparent at a long "nonlinear" time of the order $\sim \omega^{-1}(kA)^{-2}$. It has a small value due to the smallness of the wave steepness $kA$, and, thus, it takes time to manifest the effect of nonlinearity.

## *4.2. Nonlinear wave self-modulation*

Wave harmonics are dependent due to nonlinear wave-wave interactions. This interaction can lead to considerable effects accompanied by a strong energy redistribution and by phase coherence in the spectral space.

Waves over sufficiently deep water suffer from nonlinear side-band instability (alternatively, modulational or Benjamin – Feir instability), which is well-known in many fields of nonlinear physics. If a uniform wave with frequency $\omega_0$ has a weak amplitude modulation, the perturbation at the first stage exponentially grows due to the energy exchange between the carrier wave and the sidebands.

Fig. 7 displays the spectral picture of this process. The carrier wave is represented by the intense spectral peak at $\omega_0$, and the sidebands due to the wave modulation have smaller amplitudes. Only sufficiently long perturbations may be unstable. The blue thin curves show the instability growth rate, which is different for different lengths of the perturbation. The sidebands accrue energy from the carrier wave, what results in the occurrence of large waves.

This effect can be described within the weakly nonlinear weakly dispersive framework. Decomposing the deep-water dispersion relation around some dominant wavenumber, $k_0$, and combining this relation with the formula for the nonlinear frequency correction, the following relation takes place,

$$\omega - \omega_0 \approx C_{gr}(k - k_0) - \frac{\omega_0}{8}\frac{(k-k_0)^2}{k_0^2} + \frac{\omega_0 k_0^2}{2}|A|^2 . \qquad (7)$$

The corresponding evolution equation may be obtained in a formal way after the changes $\omega - \omega_0 \to i\, \partial/\partial t$, and $k - k_0 \to -i\, \partial/\partial x$:

$$-i\left(\frac{\partial A}{\partial t} + C_{gr}\frac{\partial A}{\partial x}\right) + \frac{\omega_0}{8k_0^2}\frac{\partial^2 A}{\partial x^2} + \frac{\omega_0 k_0^2}{2}A|A|^2 = 0 , \qquad (8)$$

which is the *nonlinear Schrödinger equation* (NLS). Here $A(x, t)$ − is the complex wave amplitude.

The instability starts if the wave amplitude is high enough, and the perturbation is sufficiently long (the spectral satellites are close to the mean frequency, see Fig. 7, thus the spectrum is narrow). Since the feasible amplitude of real sea waves is limited, a minimum length of physically unstable perturbations exists. Typical sea wave trains over deep water can be unstable if contain more than about 5–10 waves.

The modulational instability in physics is usually studied with respect to deterministic wave packets. But the specificity of wind waves is that they are random. The theory of modulational instability for narrow-banded random waves was developed in [Alber, 1978] in seventies, who demonstrated that the wave randomness suppresses the instability greatly. This happens when the correlation length becomes shorter than the length of modulation, then the nonlinear self-modulation supported by the wave coherence cannot occur. If waves represent a random process with the Gaussian spectrum with variance $\sigma_r^2$, then the correlation and the modulation length ratio is

$$\frac{correlation\ length}{modulation\ length} \propto \frac{k_0 A}{\sigma_r / k_0}. \qquad (9)$$

The significance of the nonlinear self-modulation effect on the stochastic wave dynamics may be estimated by the Benjamin – Feir Index, *BFI*, which is the ratio of magnitudes of the nonlinear and dispersive effects. In this case the dispersion has the physical meaning of the difference between wave velocities within wave groups. The *BFI* may be defined as

$$BFI = \frac{steepness}{spectrum\ width} = \frac{k_0 A}{\Delta\omega / \omega_0}. \qquad (10)$$

Roughly speaking, waves are unstable for *BFI* > 1 and are stable with respect to the Benjamin – Feir instability otherwise. Thus, the instability condition *BFI* > 1 is consistent with the request of the sufficient correlation length discussed just above.

For typical sea conditions the *BFI* is less or about the unity. The importance of the Benjamin – Feir instability effect for sea wind waves is now being revised in view of the rogue wave phenomenon. In particular, the sea conditions characterised by a large value of the *BFI* are supposed to be more dangerous, since the modulational instability may become triggered. The unstable modulational growth due to the Benjamin – Feir instability is a regular mechanism of a high wave generation over deep water, which increases the probability of high waves beyond the prediction of the linear theory.

### *4.3. Breather solutions*
The NLS equation (8) admits exact solutions, which describe the development of the Benjamin – Feir instability. Such an example is given in Fig. 8. The wave envelope |*A*| is shown in scaled variables in this figure; the solution is symmetric with respect to

the coordinate origin. At a large time the solution represents a weakly modulated wave train. The wave has the maximum amplitude at the time equal to zero; then it is thrice higher than the wave at the infinite time.

This solution corresponds to the infinitively long perturbation of the wave train. For a given wave amplitude there is a certain domain of perturbation wavelengths, when deep-water uniform wave trains are unstable with respect to the modulational instability, see Fig. 7. Finite but sufficiently long wave perturbations lead to smaller wave amplification than it is provided by the solution in Fig. 8, see [Osborne, 2010]. According to the NLS theory, the most unstable perturbation (with the maximum growth rate) results in about 2.4 amplification of the wave amplitude.

When many unstable modes of the wave field are excited, different growth rates and starting conditions result in a tangled competition between unstable modes. The evolution of modulationaly unstable wave trains looks quite intricate, and in real applications this dynamics is chaotic [Ablowitz et al, 2000].

The exact solutions of the NLS equation, similar to the one shown in Fig. 8, describe the huge wave occurrence 'out of nowhere', and are often called breathing waves or *breathers*. They may be described and qualitatively understood by virtue of the Inverse Scattering Technique. A rich family of solutions (multi-breathers) of this kind has been discovered recently by mathematicians in application to the rogue wave problem. The breathing solutions are now considered as prototypes of rogue waves in many branches of physics (see, for example [Akhmediev & Pelinovsky, 2010]).

A nonlinear superposition of breather waves may result in further wave intensification (multi-breather waves). An example is given in Fig. 9, where a third-order rational solution describes the 7-times wave amplification, and this is not the limit. The analytical solutions are generally rather difficult for analysis.

The modulational instability of deep-water surface waves was first observed in a laboratory tank by Benjamin and Feir. The effect of generation of very high waves from initially weakly perturbed wave trains has been confirmed many times within the recent years by means of numerical simulations of the primitive hydrodynamic equations. Such example is given in Fig. 10, where a weakly perturbed wave train is simulated in time by means of a 3D strongly nonlinear solver of the primitive equations of hydrodynamics. The wave shown in Fig. 10 is now considered by many researchers as a typical rogue wave, responsible for the abnormally frequent observation of extremely high waves.

*4.4. Coherent wave groups*

The effect of the modulational or Benjamin – Feir instability results in splitting of regular waves into wave groups. In fact, the nonlinear wave patterns over deep water can be long-living; one can see wave groups in Fig. 1a in contrast to Fig. 1b.

Nonlinear wave groups differ from ordinary, linear, wave groups in the phase coherence between wave harmonics caused by the nonlinear wave-wave interaction. A linear wave group would quickly disintegrate due to the dispersion effect, while the nonlinear wave group can remain coupled for a while. Having a look at the NLS equation (8) or, even better, at the formula for wave frequency with dispersion and nonlinear corrections (7), one may qualitatively understand the reason why nonlinear wave groups can exist: it happens when the nonlinear and the dispersive terms compensate each other.

The nonlinear Schrödinger equation is capable to describe such nonlinear wave packets, called *envelope solitons*. Envelope solitons represent the long-time asymptotic solution of the initial-value problem for the NLS equation, and, thus, have a fundamental importance. One may say that the process of the modulational instability of perturbed wave trains in the long run results in creating envelope solitary waves; the breathing waves discussed in the previous section describe the superposition of coherent wave groups with background quasi-linear waves.

Envelope solitons of the NLS equation interact elastically with other waves, and therefore, completely preserve their energy.

The breathing waves and nonlinear solitary wave groups may be detected within the stochastic sea waves with the help of the Inverse Scattering Technique (the matter of the IST may be discovered with the textbook [Drazin & Johnson, 1996], and for the application to the rogue wave detecting see monographs [Pelinovsky & Kharif, 2008; Kharif et al, 2009; Osborne, 2010]), what could be used for the early warning of the rogue wave hazard. This is a very interesting application of a mathematical approach to practically important needs.

Rogue events often represent *rogue wave groups* ('three sisters'), and sometimes the events may be reasonably well explained by wave group nonlinear dynamics. Recently, very short wave groups with very steep waves have been demonstrated by means of fully nonlinear simulations to be long-living and stable with respect to some kinds of collisions with other waves. Thus, the presence of an intense solitary wave packet may give a hint of a nearing rogue wave danger.

Solitary envelopes exist under the assumption of unidirectional waves. The situation becomes more complicated when crested sea states are considered (the 3D case, when surface waves propagate under different angles). Then the envelope solitons are transversally unstable. This is due to the fact that weak perturbations in the transverse wave direction result in the increase in the discrepancy, there is no physical effect which would relax the deviation. As a result, the problem cannot be integrated by means of the Inverse Scattering Technique, eternal solitons do not exist, and wave dynamics becomes much more complicated.

However, if the angle wave spectrum is relatively narrow, coherent wave patterns may still be observed and determine the wave dynamics, though for a shorter time scale. In Fig. 11 a result of a strongly nonlinear numerical simulation of the primitive equations of hydrodynamics is shown, when a weakly perturbed 3D wave train with a very small steepness 0.07 develops a huge breaking wave. This effect is a manifestation of the Benjamin – Feir instability, which in the 3D case results in a more sophisticated dynamics of the wave envelope.

Although Fig. 10 and Fig. 11 may look similar, the transverse dynamics in Fig. 10 has no importance; it just slightly distorts the wave shapes. On the contrary, the dynamics in Fig. 11 is essentially three-dimensional, what enables providing a much more significant wave enhancement.

It has been shown recently that the nonlinear self-modulational effect of waves over deep waters does not change noticeably the frequency of the rogue wave occurrence, if the waves are significantly short-crested. The effect is apparently due to the interplay between the 3D nonlinear modulational instabilities of water waves; though this problem has not been completely explored yet. On the other hand, there is a principal distinction between unidirectional and directional waves due to different conditions of allowed nonlinear wave resonances.

In-situ measurements of rogue wave surfaces are almost absent; studies on the relation between rogue wave statistics and the angle spectrum width are limited; laboratory experiments in shaped wave tanks are complicated and costly. Thus, now numerical simulations build up the main basis for studies of the 3D rogue wave dynamics, and are intensively carried out.

## 5. Nonlinear shallow-water rogue waves

Shallow-water waves exhibit rather different properties in comparison to the deep-water ones. Their shape in a very shallow water is much more asymmetrical with respect to the mean sea level, than the shape of deep-water waves, see Fig. 1. Shallow-water waves possess a weak dispersion, what leads to relatively long 'life-time' of individual waves.

Examples of rogue waves, which were recorded in shallow water conditions of the Baltic Sea at the 2.7-meter depth, are displayed in Fig. 12. There is a variety of shapes of shallow-water rogue waves. The wave grouping is not so appreciable, and the rogue waves are often single waves.

It has been already pointed out that when wind waves propagate over the shallow water, they are strongly influenced by peculiarities of the sea floor bathymetry, which varies significantly. The effects due to the topography variation and when waves approach the shoreline will be discussed in more detail later on, in the section devoted to *coastal effects*. At first, we consider water waves in a basin of constant depth.

The wave velocity is determined by the dimensionless parameter *kh*, which is small in the case of shallow water; it is altered due to the dispersion. The steepness, which characterizes the nonlinearity of deep-water waves, is now not an appropriate characteristics (at least, outside the surf zone); instead, the shallow-water wave nonlinearity is controlled by the ratio

$$\frac{wave\quad amplitude}{water\quad depth}. \qquad (11)$$

The ratio of the nonlinear parameter over the dispersion estimator composes the *Ursell number*, *Ur*, which characterizes the significance of the nonlinear dynamics of shallow water waves:

$$Ur = \frac{(wave\ amplitude)(wave\ length)^2}{(water\ depth)^3}. \quad (12)$$

It plays the same role as the BFI number in the deep-water case.

*5.1 Steepening of shallow-water waves*

When the Ursell parameter is very large ($Ur >> 1$) then the nonlinearity prevails, while dispersion effects may be ignored at the first approximation. Thus, waves undergo steepening, becoming more asymmetric in terms of the face-back slope asymmetry. The brightest example of this fact is the surf wave steepening and plunging near the coast.

The shallow water equations, when disregarding the dispersion, are nonlinear hyperbolic equations which are completely similar to the ones used in nonlinear acoustics, and gas and magneto-hydrodynamics.

Then, the solutions describing nonlinearly deforming waves are called *Riemann waves*. Riemann waves propagating over the shallow water are shown in Fig. 13 in a simplified manner. It can be seen that the wave shape changes towards more asymmetrical during the propagation, while the amplitude of the wave remains same, until the wave overturns. During this process the wave energy is transferred to shorter scales. The process of wave steepening can be explained by the difference in wave speeds under the wave crest and under the trough (the water column is larger under the wave crest, rather than under the trough). Hence, the wave crest propagates with a greater velocity and overtakes the trough forming a steep front.

Under the assumptions implied, this effect may be described analytically, and the wave breaking phenomenon corresponds to the infinite value of the wave slope steepness (in mathematical terms it is called a gradient catastrophe in hyperbolic equations). The intense wave energy transfer to shorter scales results in a typical for shallow water wideband spectrum; the strong wave asymmetry is supported by the strong phase correlation of the wave components.

Similarities with the nonlinear acoustics and magneto-hydrodynamics do not extend further. The water wave breaking effect has its own peculiarities. At the breaking wave crest hydrodynamic instability phenomena become crucial, they lead to the fluid turbulization and the growing importance of the turbulent viscosity effects. These processes are most efficient for large waves ($H > 1.5\,h$) and result in appearance of so-called hydraulic jumps (equivalent to shock waves in the compressible gas).

In the case of a relatively small-amplitude wave ($H < 1.5\,h$) the dispersive effects become important due to the steep wave front, and are able to scatter different spectral components corresponding to different wave scales. This leads to the formation of a wavy structure on the shock wave front, the so-caller *undular bore*. Mathematically, it is described by Boussinesq-type systems, presenting a dispersive generalization of the shallow-water equations. The most spectacular tidal bores can be observed in rivers Severn (UK), Seine (France), Qiantang (China) and in the Turnagain arm of Cook Inlet (Alaska, USA), see [Chanson, 2011].

*5.2. Cnoidal and solitary waves*

When the weakly nonlinear and weakly dispersive effects are both taken into account, what corresponds to the values of the Ursell parameter of the order of unity, shallow water waves in many cases can be sufficiently well described by the Korteweg – de Vries equation

$$\frac{\partial \eta}{\partial t} + C\left(1 + \frac{3\eta}{2h}\right)\frac{\partial \eta}{\partial x} + \frac{Ch^2}{6}\frac{\partial^3 \eta}{\partial x^3} = 0. \qquad (13)$$

Here $\eta(x, t)$ is the water surface elevation.

The steady-state solution of the Korteweg – de Vries equation describes travelling waves of a permanent form – *cnoidal waves*, named after the Jacobian elliptic function *cn*. Such waves play the same role in the shallow-water dynamics as the Stokes waves, previously described in Section 4, in the deep-water case. A steep cnoidal wave is shown in Fig. 14 in comparison with a sinusoidal wave of the same height. While almost sinusoidal in the small-amplitude limit, the intense cnoidal waves resemble a train of isolated wave humps – *solitons*.

Being originally discovered in water, nowadays solitons play a fundamental role in the modern nonlinear physics. A surface water soliton was first described in 1844 by John Scott Russell. He observed it in 1834, when a sudden stop of a boat, that was moving with small acceleration in a channel of uniform depth, preceded the occurrence of an "exotic water dome" or "wave of translation", called later a *soliton* due to its unique similarity to a particle. Later on, the existence of solitons was confirmed mathematically through the obtaining of steady solutions of the Boussinesq and Korteweg – de Vries equations. Indeed, solitons may propagate for a long distance without energy loss, interacting elastically between each other and with other waves.

Solitons have been discovered in many other important physical equations (including the envelope solitons of the nonlinear Schrödinger equation, discussed above), but the *first* soliton was observed on the shallow water surface. Groups of solitary humps are often well-seen on photos of undular bores, mentioned above [Chanson, 2011].

### *5.3. Focusing mechanisms in shallow water of constant depth*
Efficient for deep-water waves modulational instability is not effective over the shallow water, and uniform waves are stable. However, nonlinear wave-wave interactions can be very important for the shallow water wave dynamics. Focusing mechanisms taking into account the wave nonlinearity are considered below. These mechanisms are manifested differently in the 2D (unidirectional wave propagation) and 3D geometries.

*5.3.1. Unidirectional fields*

It is convenient to consider nonlinear wave-wave interactions over shallow water with the help of exact solutions of the Korteweg – de Vries equation, which describe collisions of soliton waves. First of all, intense solitons may not be called unexpected rogue waves due to the fact that they are stable and long-living. When Korteweg – de Vries solitons interact, the maximum wave amplitude does not exceed the amplitude of the largest soliton, in contrast to the linear superposition of two waves. Therefore, co-moving solitons by their own do not spawn rogue events.

At a first glance this fact may look surprising, that the result of interaction of weakly nonlinear waves (they may be almost linear waves) is far from the prediction of the linear superposition. However, solitons are maintained by the balance between the dispersion and nonlinear effects, which are supposed to be of the same order. Therefore, when the nonlinearity of solitary waves is neglected, then the dispersion should be disregarded as well, and thus two solitons will travel with equal velocities, and, consequently, will not collide. Concluding, the soliton interaction is essentially related to the wave nonlinearity and does not have an analogue in the linear theory.

The temporal focusing of shallow water waves due to the dispersion is still possible, since linear short-scale waves have different velocities. The dispersion focusing is efficient if a large number of wave components is involved; all together they are able to compose a huge wave. The presence of solitary waves, their interaction between each other and with other wave trains do not cancel the dispersion focusing effect. Solitons participate in the wave focusing process, providing variety of rogue wave shapes.

The probability of rogue wave occurrence within the Korteweg – de Vries approach has been studied by numerical simulations of irregular wave fields produced with the spectra similar to ones measured in the coastal zone of the North Sea; these conditions correspond to shallow-water wave spectra. It has been demonstrated that the increase in the Ursell parameter leads to the growth of the third statistical moment for the water surface elevation, meaning that intense crests prevail in the wave field when compared with troughs. A non-monotonical behaviour of the fourth statistical moment (kurtosis) is also shown; the rogue wave probability decreases for weakly nonlinear waves and increases for strongly nonlinear ones, see [Kharif et al, 2009] for details.

*5.3.2. Directional fileds: Nonlinear geometrical effects*

In contrast to the deep-water envelope solitary groups, trains of cnoidal waves and shallow-water solitons are stable with respect to transversal perturbations. Trains of planar (i.e., elongated in the lateral direction) peaked waves are frequently observed in the coastal zone (see Fig. 15).

It is evident from the physical point of view that if waves propagate under a large angle between each other (including the case of opposite-directed waves), their interaction is weak and the waves superpose almost linearly. This process is actually the linear geometric focusing, and can produce a rogue event in the focus. When nonlinear effects are taken into account, they slightly modify features of the focusing process and the focussed wave.

A new effect occurs when the waves propagate under a small angle between their directions. In this case the nonlinear three-wave resonance condition can be satisfied, when the third wave, which propagates in an almost perpendicular direction, should be accounted for. This effect may take place in media with no dispersion (like the compressible gas dynamics), and with a weak dispersion (the shallow water case), and is known as the formation of the *Mach stem*. This kind of the essentially nonlinear wave interaction can lead to the generation of large pulses. The rogue character of the intense waves generated through this effect may be addressed to the weak variation of the angle between the interacting waves.

The Mach stem can be explained by considering the interaction of two oblique propagating solitary waves within the framework of the Kadomtsev – Petviashvili equation, which is a generalization of the Korteweg – de Vries equation for the case of a weak transverse wave variation. As it is said just above, the appearance of the wave collision area strongly depends on the angle between the directions of the soliton propagation. The Kadomtsev – Petviashvili framework reports that this effect may result in 4-times wave amplification, and 8-times wave slope increase, see Fig. 16.

*5.4. Generation of rogue waves nearshore*

When wind waves approach the sea shore, they are influenced by the sea bathymetry. Expression (3) tells that waves propagate slower when the sea becomes shallower; then the role of nonlinear effects increases, according to formulas (11) and (12).

Strong bathymetry variations result in wave transformation, refraction, diffraction and reflection what has been pointed out in the introduction; to a certain extent the effects can be described within the linear approximation. As a result, waves may converge and form areas of wave intensification, where large waves are more frequent. In such areas rogue waves can occur, as it is discussed in section 3 in relation to the 'linear' rogue waves. The nonlinearity, of course, influences this process changing characteristics of rogue waves (the height, shape), for instance, due to the breaking of high-amplitude waves.

*5.4.1. Edge wave dynamics*

A new effect related to the nonlinearity and the bottom relief variability is the generation of rogue *edge waves* in the coastal zone. Edge waves are a kind of waves trapped by the bottom topography, which can form wave guides. In the simplest case of the uniform beach along a straight coastal line, edge waves propagate alongshore. In the offshore direction the waves are described by the eigenfunction of the corresponding Sturm – Liouville problem with the boundary condition decaying exponentially towards the open sea.

Edge waves are often considered as the major factor of the long-term evolution of the coastal line, forming the rhythmic crescentic bars, similar to the ones shown in Fig. 17. In the linear theory, there is a set of independent modes of the edge waves with the dispersion relation $\omega_n = \sqrt{n\alpha g k}$, where $\alpha$ is the bottom slope, $k$ is the alongshore wavenumber, and $n$ is the eigenmode number.

As it can be seen from the dispersion relation, these waves are strongly dispersive, similar to the deep-water waves case described in section 4. Moreover, nonlinear theory reports that edge waves are modulationaly unstable, and therefore the nonlinear dynamics of edge waves may be quite similar to the nonlinear dynamics of deep-water wind waves. Therefore, the potential existence of edge rogue waves is evident; their properties are described in [Akhmediev & Pelinovsky, 2010].

It is important to mention that edge waves are usually generated through the nonlinear interaction between wind waves moving in almost onshore direction. The typical frequency of edge waves is about half of the typical frequency of the wind waves; thus, the typical period of edge waves is about 10–20 s. The characteristic wavelength of edge waves (in the alongshore direction) is *4nα* times the wind wave length, what gives typically 50–300 m. In the offshore direction edge waves have approximately the same length, and therefore, edge rogue waves can form isolated bells of surface elevation (or groups of waves of different polarities) with the diameter of about 50–300 meters.

*5.4.2. Wave interaction with steep coasts*
Another kind of extreme events in the coastal zone is related to the strong wave interaction with the coast. In the first approximation the simplest geometry of the coastal zone may be analyzed, when a basin of constant depth is bounded by a straight vertical wall, what models steep cliffs or seawalls. When the frontal approaching of a small-amplitude wave is concerned, and the wave is neither very asymmetric nor breaking, the effect of nonlinear interaction between the incident and reflected waves is weak due to the short time of their collision. Therefore, this kind of wave-coast interaction represents an almost linear superposition of the incident and reflected waves. The nonlinearity changes the probability of freak wave occurrence increasing the probability of high wave crests, but this effect is not so strong.

A more dramatic effect is caused by glancing waves approaching the coast under a large angle (close to 90°) to the onshore direction. Since the boundary condition on the wall retains the mirror symmetry, this case is quite similar to the case of weakly crested waves, which has been described above in the framework of the Kadomtsev – Petviashvili equation. Therefore, the Mach stem can be observed near walls as well. The height of the Mach stem can be 4-times larger than the height of the incident wave, and the wave steepness may be enhanced in 8 times.

If the incident wave has an asymmetric shape due to nonlinear effects or due to the overturning process, ultimate wave amplification can be achieved. In this case the pressure at the steep wave front ceases to obey the hydrostatic assumption. As a result, the kinetic energy of the incident wave transfers into the kinetic energy of the vertical water motion (which is usually small in the hydrostatic approximation) forming a jet and a splash at a large height.

When the approaching wave is almost broken, it can capture a big volume of air, and then the air bubble comes to the surface and gives an additional upward impulse to the water. Nowadays such effects are being modeled within the framework of fully nonlinear hydrodynamic Euler or Navier–Stokes equations. High splashes are often observed at cliffs or steep sea coasts (see a photo in Fig. 18). The rogue character of this phenomenon is related to the high sensitivity of the splash height with respect to the incident wave amplitude and steepness.

The effect of vertical cumulative jets may have great relevance to another aspect of rogue wave problem. These jets are able to cause damage to offshore stationary oil and gas platforms. When sea waves interact with the platform base or its legs, they can splash upward. Since this effect is strongly nonlinear, even slightly higher waves can produce a significantly stronger effect on the platform deck. Mentions about similar accidents exist, but the information about them is generally poorly documented.

*5.4.3. Gentle beach flooding*
Observations of rogue wave events at coasts are of a special interest and may be emphasized. Because at all times coastal areas have naturally been inhabited by people, there are numerous eye-witness observations, descriptions and video recordings which imprint rogue wave occurrence near and on the shoreline. These rogue wave events usually represent short-time sudden floodings of dry beach or giant splashes onto and over protecting embankments.

It is significant to mention that heavy floodings are commonly associated with tsunami wave runups and storm surges. However, a large amount of data is known when a beach flooding failed to be related to any generation sources. A description of this kind of event is reproduced from [Bryant, 2008]:

> "One of the more unusual events took place on the island of Majuro in the Marshall Islands in 1979. On a clear calm day, a single 6 m high wave appeared from the northeast at low tide, crossed the reefs protecting the shoreline, and crashed through the residential and business districts in the town of Rita, washing away 144 homes. The next day at high tide the same thing happened again. After this second wave hit, the island was declared a natural disaster area by the U.S. government, which was administrating the islands. Six days later, another series of waves up to 8 m high again swept the east coast of the island, destroying the hospital, communications center and more houses. The waves cost $20 million and affected the lifelihood of two-thirds of the island's 12,000 people."

A testimony of an incident, when a sudden flooding of the coast occurred in Mavericks Beach (California, USA) on 13 February 2010 is displayed in Fig. 19. Today, the effect of a sudden flooding due to unknown reasons is referred to the rogue wave problem in the context of the near shore wave dynamics, and is under investigation.

The theory of rogue wave phenomena on gentle beaches is not as much developed as for the case of waves in water basins. However, some mechanisms of the anomalous wave amplification can be pointed out.

First of all, a gentle slope of the nearshore bottom causes a substantial wave change while it is approaching the shore. When the water depth decreases, waves propagate with a lower velocity as has been already discussed. Due to the smallness of the bottom slope, the wave reflection may be ignored and, therefore, the wave energy flux is conserved. Keeping in mind that the wave energy is proportional to the squared wave amplitude, $A^2$ (the coefficient of proportionality includes the gravity acceleration and the water density), and the speed of wave propagation is defined by the shallow-water relation, $C = \sqrt{gh}$, the Green's law $A \sim h^{-1/4}$ follows from the energy flux conservation law, confirming that the wave amplitude grows when approaching the coast. This process of the wave amplification near the coast is called *shoaling*.

The simple explanation given above makes clear, why waves in the coastal zone become higher and steeper. The process of wave transformation when climbing up the coast strongly depends on the shoaling depth profile. Some of the profiles provide conditions for an efficient transport of wave energy towards the coast. Generally, gentle coasts are more favorable for efficient wave energy transfer. Abrupt depth changes result in wave refraction and thus may reflect wave energy preventing its passage onshore.

The energy transmission ability depends also on the wave frequency: short waves propagate according to the Green's law, but longer waves (in comparison with the beach characteristic scale) do not undergo a slow change due to the water shoaling; instead, they are reflected from the coast as from the wall. As a result, the nearshore zone represents a kind of a frequency filter with the transmission coefficient growing with the frequency. This explains why the spectrum of waves at the coast is shifted to higher frequencies.

Another consequence of the described above effect is that the Ursell parameter increases near the coast and nonlinear effects prevail; they lead to the strong deformation of the wave shape. The steep wave crests are even stronger amplified than smoother wave troughs. Wave troughs touch the ground and reflect from the coast before the wave crests do; the latter pass along the dry beach causing its flooding. Observations confirm the qualitative conclusions that the average sea level is raised (the set-up); the statistical distribution of the water level deviates from the Gaussian one.

The surf zone, where waves transform into the turbulent flow is generally difficult for a theoretical analysis. The runup characteristics can be evaluated within the framework of the nonlinear shallow-water theory under the assumption of non-breaking waves. The vertical oscillations of moving shoreline in the linear (dashed line) and the appropriate nonlinear (solid line) theories are shown in Fig. 20. It can be seen that nonlinearity does not change positions of the extremes, but increases the duration of the wave flooding and decreases the duration of the backwash. Consequently, the nonlinearity does not affect the distribution of the runup amplitudes, but influences the overall statistics of the moving shoreline and, in particular, leads to the mean sea level rise (the wave set-up), which becomes more significant for a stronger nonlinear case.

The described above examples demonstrate how the competition of nonlinear and dispersive effects, the bottom and coastal line variability make the rogue event appearance at the coastal line manifold and difficult for the interpretation and analysis. In many cases the nonlinearity plays an important role amplifying the extreme character of the wave dynamics and providing a specific rogue wave statistics.

**Conclusion**
In this paper the present state of the rogue or freak wave problem is overviewed from a very popular and physically general point of view. This problem has been recently recognized, and the most significant actual trend of its extension is directed towards the consideration of nonlinear and strongly nonlinear effects in the water wave dynamics. In this paper the main suggested effects enabling rogue wave generation are listed and briefly explained. The deep-water, shallow-water and coastal areas have been considered, emphasising the difference between the water wave physics at these conditions. For more details see reviews [Kharif & Pelinovsky, 2003; Dysthe et al, 2008] and monographs [Kharif et al, 2009; Osborne 2010].

Many mechanisms underlying the rogue wave phenomenon have quite general physical nature and may be addressed to many other applications in solid matters, superconductors, plasmas and nonlinear optics, and even financial theories. The boom of the research in oceanic rogue waves has already spawned studies in other fields. In particular, optical rogue waves may be very well described by the nonlinear Schrödinger equation, and thus, are modelled by the discussed analytical solutions with minimal modifications [Kibler et al, 2010]. The superfluid helium turns out to be a convenient model for studying the nonlinear wave turbulence in laboratory conditions [Ganshin, et al, 2008; Efimov et al, 2010].

Another exciting feature of the rogue wave study is that the modulational instability effect has been reconsidered. The modulational instability was discovered before the elaboration of the Inverse Scattering Technique (IST), and many classic results in optics and other media related to the modulational and self-modulational instabilities were obtained by virtue of approximate methods. Now these results are at the new level of understanding. Furthermore, the very mathematical IST approach is now being applied to practical needs of detecting the 'unstable' modes containing the growing oceanic billows.

Since the dynamics of real sea waves is defined by many various phenomena, related to different spatial and temporal scales, often stochastic and uncertain, the solution of the rogue wave problem cannot be limited by theoretical studies, but should involve laboratory and natural observations. Due to the recent development of the computing machinery, laboratory facilities and measuring equipment, the most realistic study of rogue wave phenomenon becomes possible and is going on.

Acknowledgements

The authors are grateful to Zh. Cherneva, S. Haver and C. Guedes Soares for provided in-situ measurement data, and to S. Anderson and R. Brander for provided photographs.

Different parts of the research have received funding from the European Community's Seventh Framework Programme FP7-SST-2008-RTD-1 under grant agreement No 234175 and RFBR grants 11-02-00483, 11-05-00216 and 11-05-92002. AS acknowledges the support from the EC Programme FP7-PEOPLE-2009-IIF under grant agreement No 254389, and grant MK-6734.2010.5. ID acknowledges the support from the targeted financing by the Estonian Ministry of Education and Research (Grant SF0140007s11), Estonian Science Foundation (Grant ETF8870) and grant MK-4378.2011.5. EP is grateful for the support from the Volkswagen Foundation.

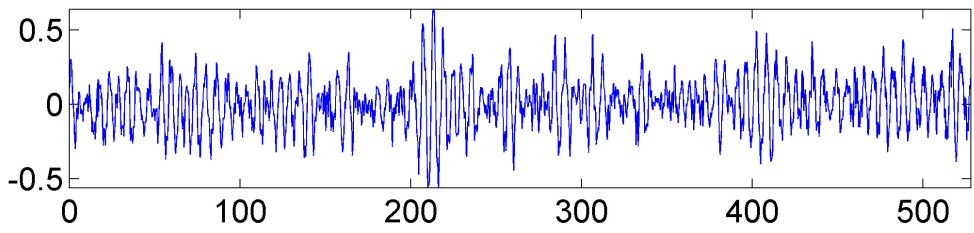

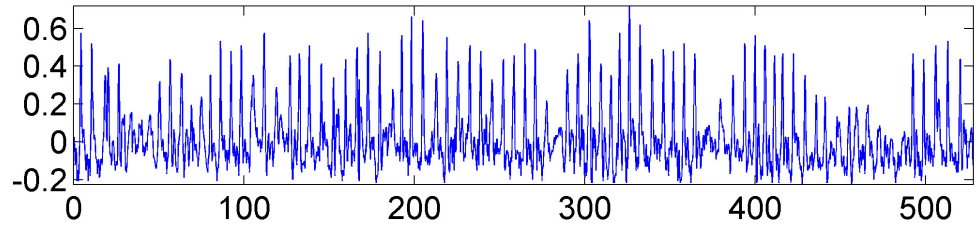

Fig. 1. Time series of the surface elevation (meters vs seconds) obtained at one coast at different water depths: 18 m (a) and 1.6 m (b). The data are granted by Zh. Cherneva, see [Cherneva & Guedes Soares, 2005].

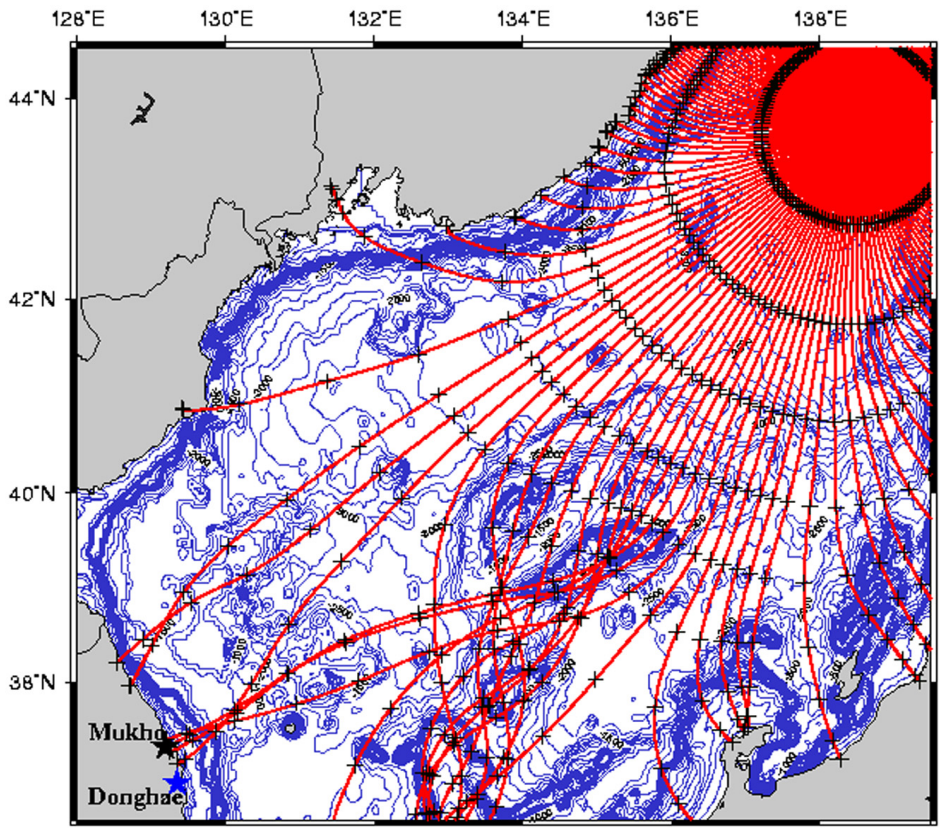

Fig. 2. Ray patterns of long waves traveling from an isotropic source in the Japan Sea. The result of numerical simulations.

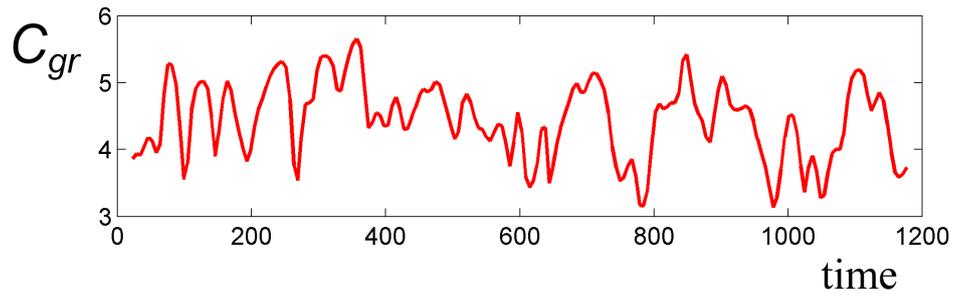

Fig. 3. An example of the local wave group velocity (in meters per second) within a 20-minute wave record retrieved in the North Sea. The typical wave period is about 10 seconds.

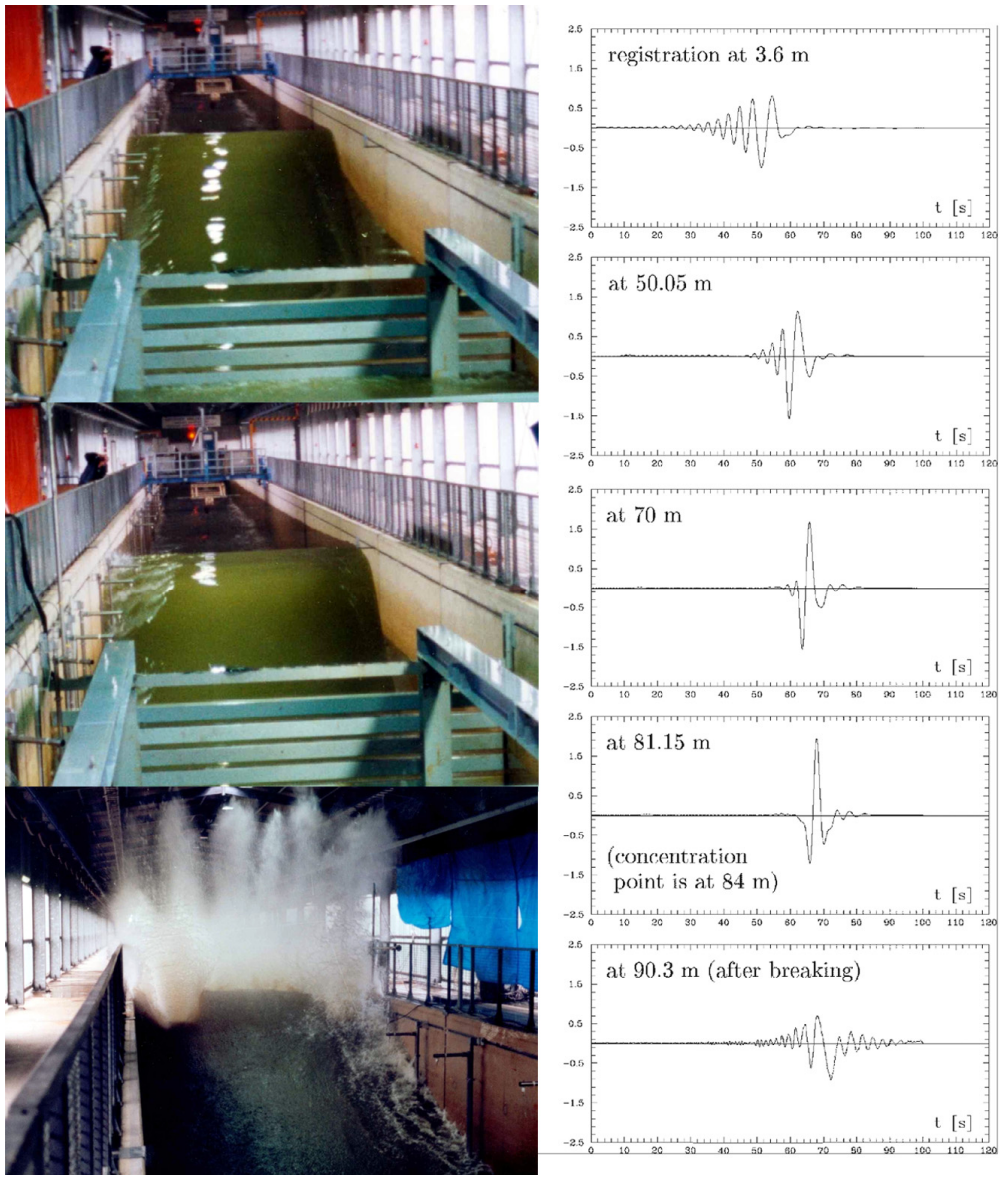

Fig. 4. The dispersion focusing of a frequency-modulated wave train in a laboratory tank. Reproduced with permission from [Clauss, 2002].

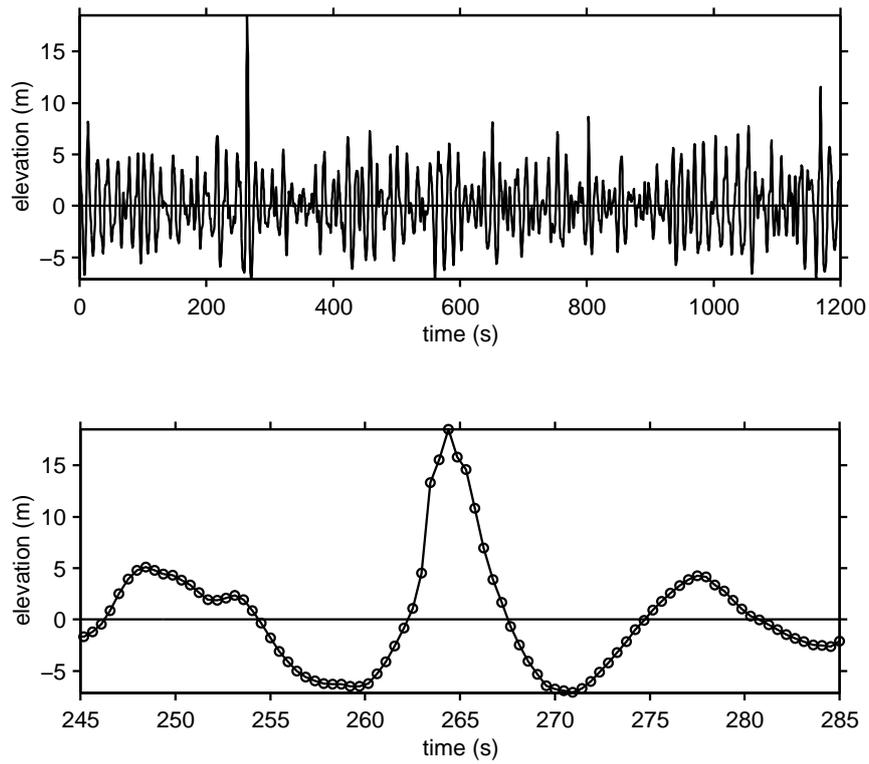

Fig. 5. Measured rogue wave time series: the "New Year Wave" (the North Sea, Draupner platform, 85 m depth, $AI = 2.24$, $H_{fr} = 26$ m). The data are granted by S. Haver.

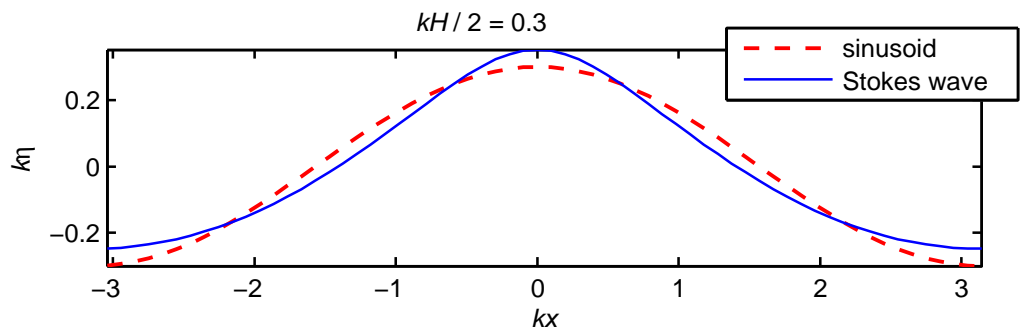

Fig. 6. Surface displacement corresponding to a sinusoidal wave and to the Stokes wave of the same height, $kH/2 = 0.3$.

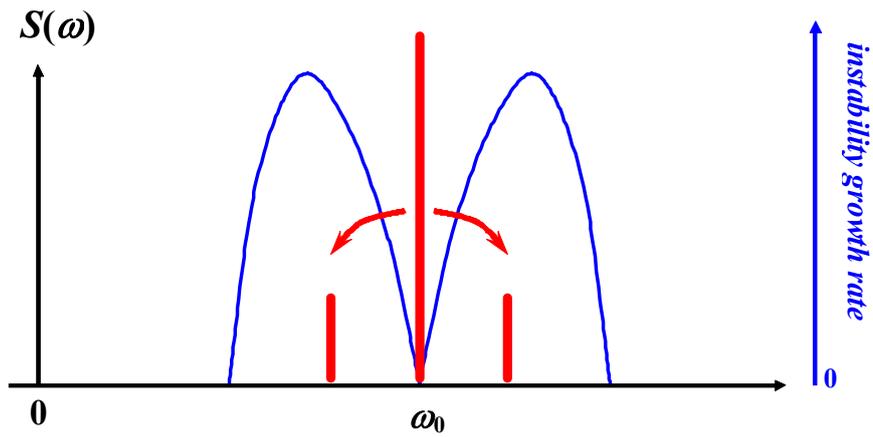

Fig. 7. Illustration of the side-band instability. The frequency spectrum $S(\omega)$ is represented by three harmonics (red thick sticks): the carrier wave at $\omega_0$ and the satellites (smaller sticks). The instability growth rate depending on the off-set frequency of the satellites is shown by the thin blue solid line. The energy transfer due to the modulational instability is shown by arrows.

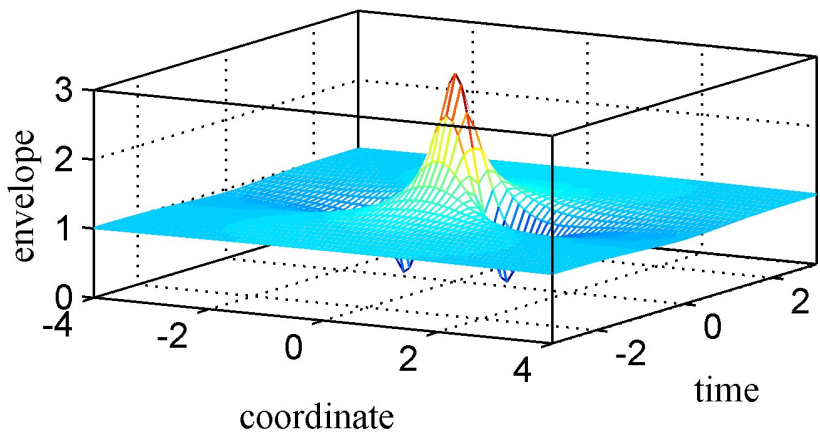

Fig. 8. An exact solution of the NLS equation, which demonstrates the occurrence of a huge wave 'out of nowhere' (the Peregrine solution).

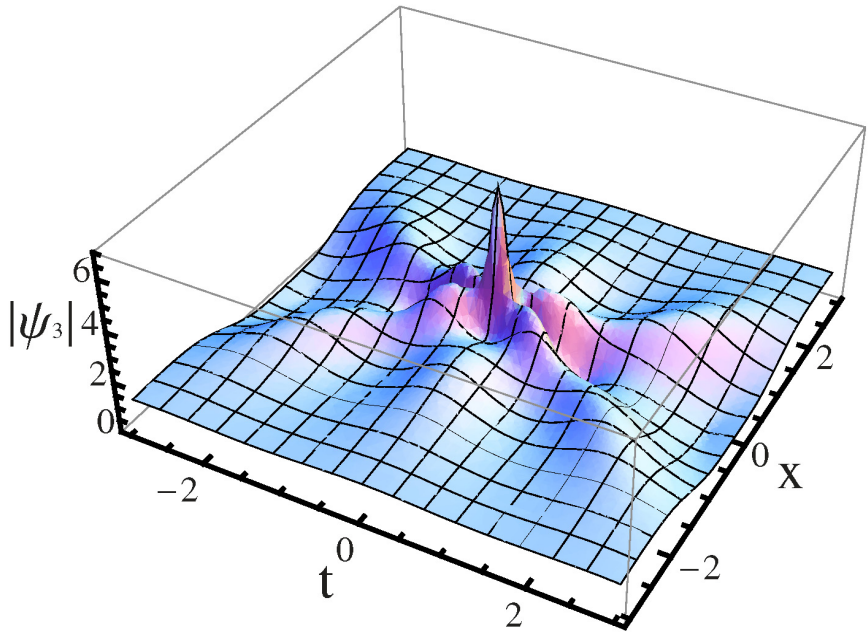

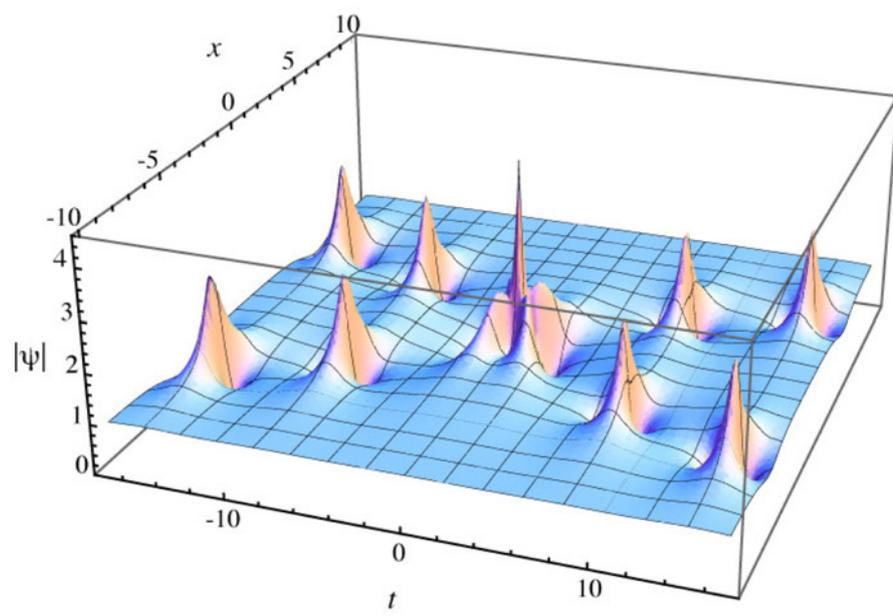

Fig. 9. Exact breather solutions of the NLS equation: a) a solution exhibiting 7-times amplification of a weakly perturbed uniform wave. Reproduced with permission from [Akhmediev et al, 2009a]; b) breather wave collision resulting in further wave enhancement. Reproduced with permission from [Akhmediev et al, 2009b].

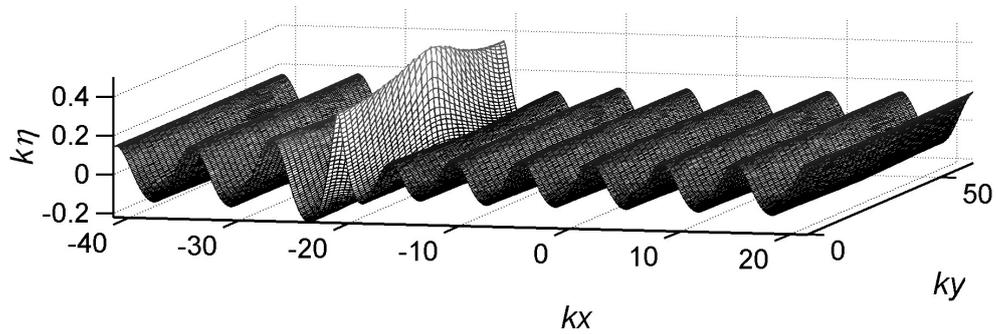

Fig. 10. A rogue wave generated due to the Benjamin – Feir instability of a weakly perturbed wave train. The initial condition has the steepness about 0.15. The resulting maximum wave overturns. The displayed moment is close to the wave breaking.

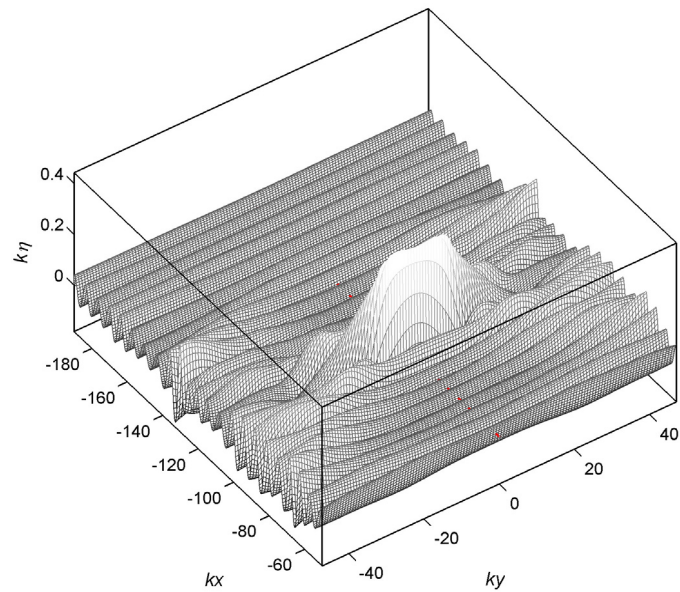

Fig. 11. A rogue wave pattern generated in the course of essentially 3D dynamics of modulationally unstable waves. The initial condition has the form of weakly perturbed wave train with steepness about 0.07. The resulting maximum wave overturns. The displayed moment is close to the wave breaking. Only a small part of the simulated surface is shown.

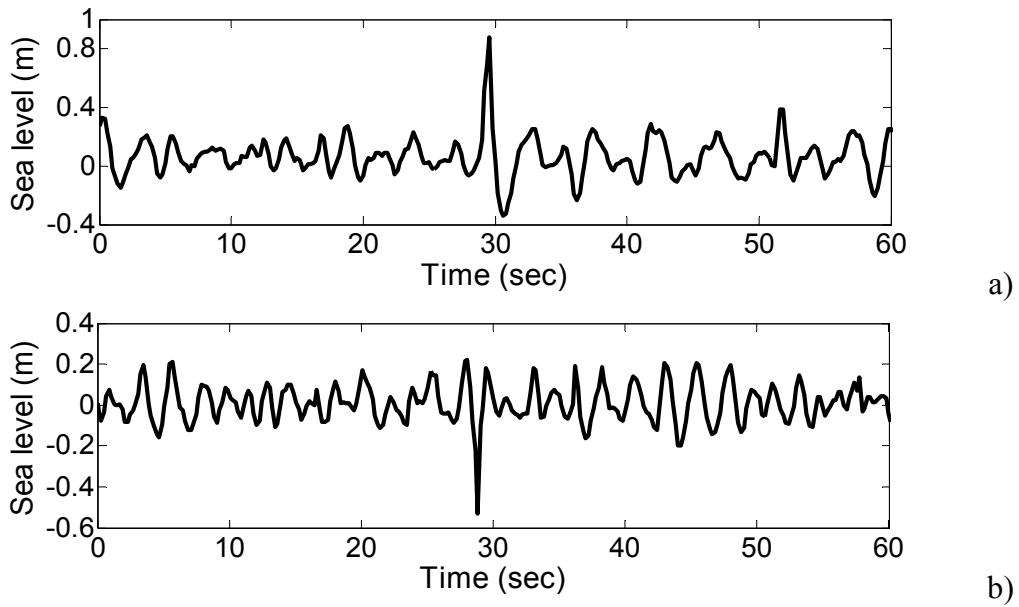

Fig. 12. Rogue wave records from the Baltic Sea measured at the 2.7-meter depth: the unexpected high wave crest (a) and the "hole in the sea" (b).

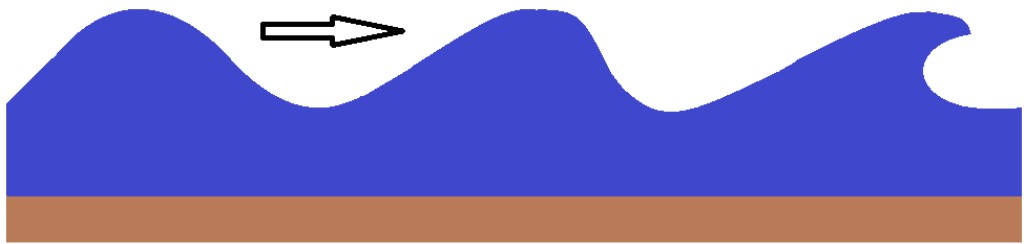

Fig. 13. Steepening of a Riemann wave during its propagation.

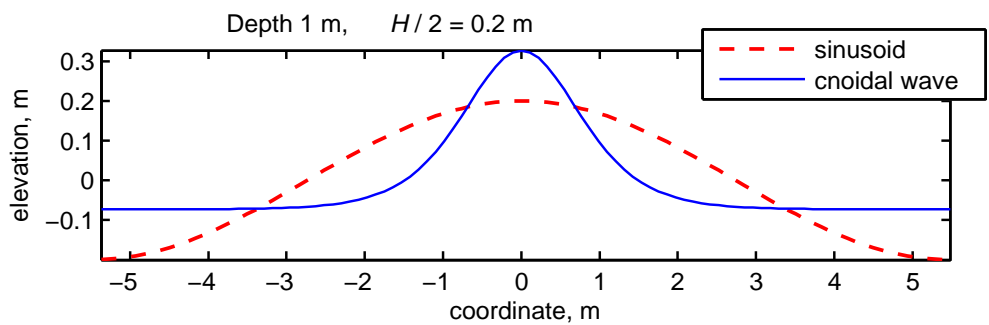

Fig. 14. A steep cnoidal wave of 40 cm height in the basin of 1 m depth, and a sinusoidal wave of the same height.

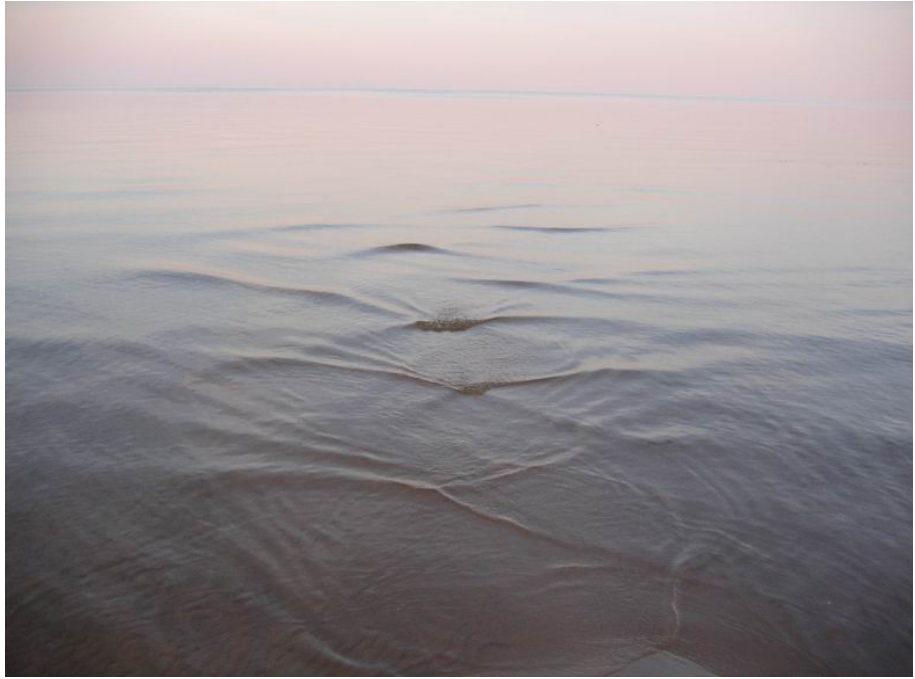

Fig. 15. A rogue wave formation in the coastal zone of Peipsi Lake (Estonia).

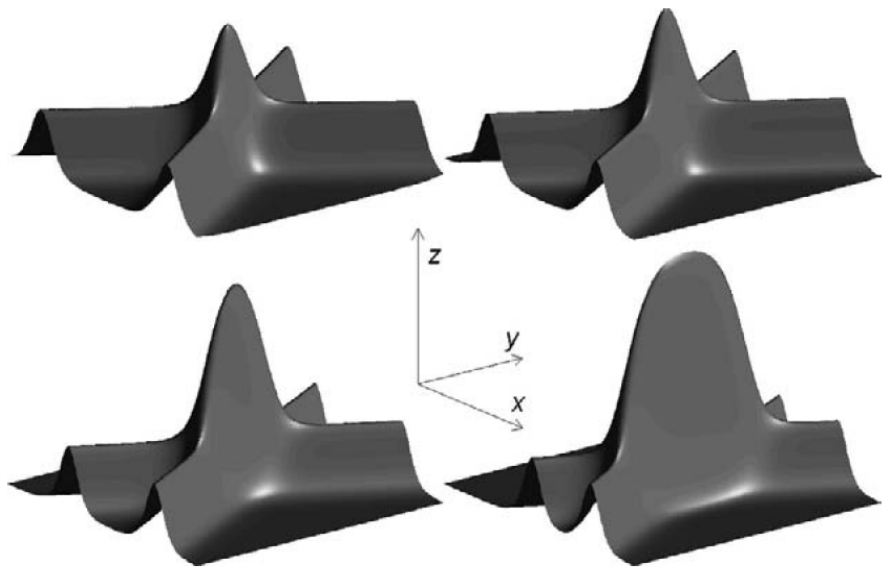

Fig. 16. Amplified waves due to the collision of planar solitary waves propagating under different angles within the Kadomtsev-Petviashvili equation framework. Reproduced with permission from [Peterson et al, 2003].

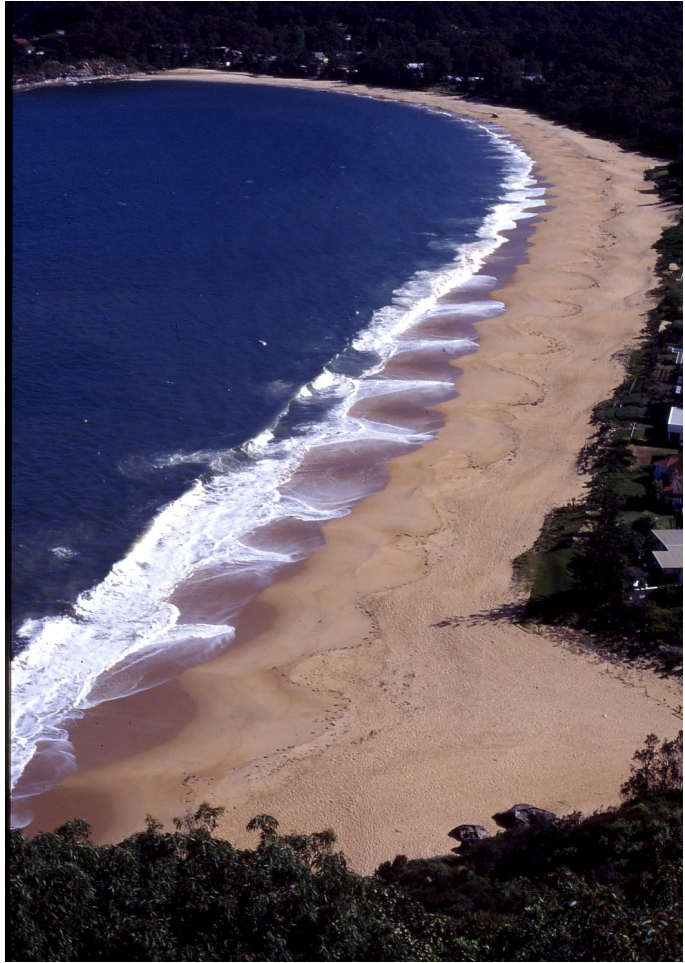

Fig. 17. The evidence of a well-defined cusp morphology and swash circulation (© Rob Brander 1996).

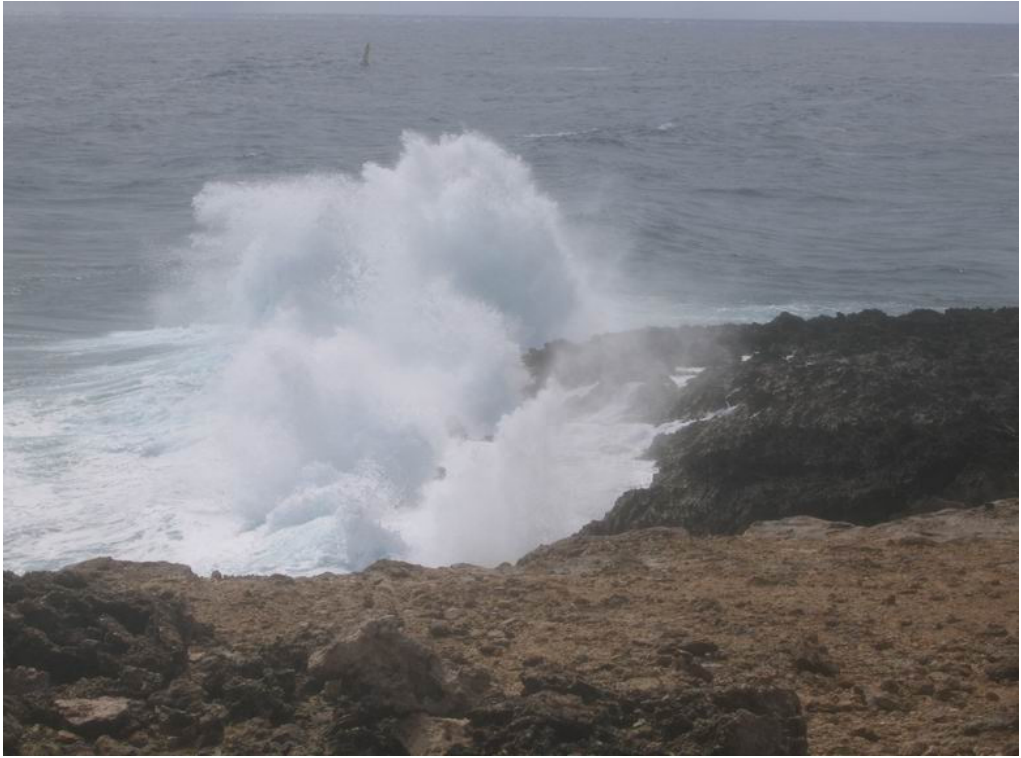

Fig. 18. A wave splash on a cliff of Petite Terre Island, Guadeloupe, France (2008).

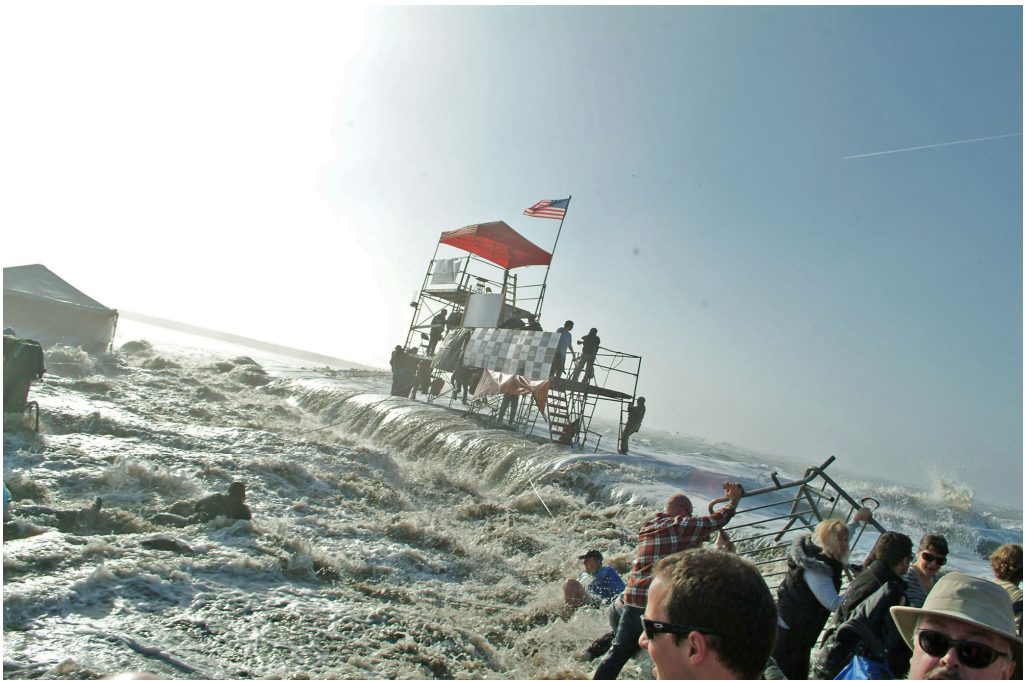

Fig. 19. Sudden flooding of Mavericks Beach (California, USA) on 13 February 2010. The water is coming to the left, overtopping the coastal wall located to the right (© Scott Anderson).

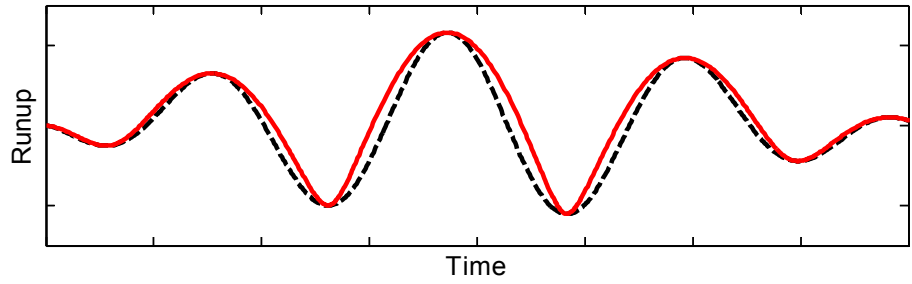

Fig. 20. Wave runup on a beach; the dashed line corresponds to the linear case (the nonlinearity is artificially disabled) and the solid line reflects the influence of the nonlinearity. Results of a computer simulation, dimensionless variables.